\documentstyle[amsfonts,amssymb,preprint,aps]{revtex}
\tightenlines
\begin{document}

\title{Graph kinematics of discrete physical
objects:\\ beyond space-time. II. Microobjects structure\\ in
two-layer matrix approximation}
\author{V. E. Asribekov}
\address{All - Russian Institute for Scientific and
Technical Information, VINITI, Moscow 125315, Russia\\
({\rm e-mail:peisv@viniti.ru})}
\maketitle

\begin{abstract}
A concrete analysis of the general properties and
numerical characteristics of different atomic and nuclear shell
systems and subnuclear particles is carried out on the base of the
solution scheme for an introduced in part~I physical graph
kinematics which give rise to the two-layer matrix representation
of the structure of any discrete
physical microobject within the self-consistent ``graph geometry''.
It is given a Riemann's foundation of the discrete manifolds in the
infinitesimal and a creation of the discrete quantity notion with
an operation of ``count'' of ``homogeneous elements'' (instead of a
``measurement'' with a scale for the continuous quantity). For
``graph microgeometry'' such special consideration in the framework of
Heisenberg's $S$-matrix particle formalism leads to the proper
``internal geometry'' with root trees basis provided for, in the first
place, an estimation of all masses of topologically different subnuclear
particles.
\end{abstract}

PACS: 11.90, 12.90, 02.10

\section*{1 Introduction}

It is noteworthy that the new physical graph kinematics in part~I
(see Ref.~[1]) is applied to the discrete physical objects and the
advantages of this formalism could be shown especially in case of
the topologically different  discrete microobjects consideration
i.~e. under transition to the ``microgeometry'' or to the space
infinitesimal.

\subsection*{1.1 Riemann's discrete manifolds in the
infinitesimal\\ and the ``graph geometry''.}

According to the classical Riemann's work of 1868 ``\"Uber die
Hypothesen, welche der Geometrie zu Grunde liegen'' (Ref.~[2]) the
metric relations principle is contained already in the discrete
manifold (die Mannigfaltigkeit) notion itself and from the
quantitative point of view a comparison of separate parts of the
discrete manifolds is carried through the ``count'' of homogeneous
elements (for the continuous manifolds~--- by means of the
``measurement'' with a scale). Under transition to the space
infinitesimal the metric relations in the ``microgeometry'', as it
is concluded in [2], are not explained by usual geometrical
admissions and the discrete quantities itself have to form
straightforwardly the discrete manifold.

A main difficulty at the creation of a notion of any discrete
mathematical or physical quantity consists therefore in the
appropriate choice of some general notion with a set of the
different ``states'' or ``determination modes'' (die
Bestimmungsweise) which must precede the creation of the discrete
quantity notion. However within the framework of the considered
graph kinematics as a such set of the different ``states'' or
``determination modes'' appears just a set of the different root
$v$-trees (counting homogeneous elements) into the microobject
``internal geometry''. Thus the ``internal geometry'' of
microobjects become structural and thereby it is realized the
self-consistent ``graph geometry'', in particular with the
quantitative characteristics~--- the definite numbers of
``discrete'' root $v$-trees.

Very such alternative way of an introduction of the postulated by
Riemann discrete manifolds was supported by Poincar\'e in due time
(see Einstein paper ``Nichteuklidische Geometrie in der Physik''
(Ref.~[3])) but the further development had stopped apparently
because of an absence of the suitable ``counting homogeneous
elements'' for the realization of ``internal geometry''
corresponding to the discrete physical objects. Nevertheless there
is a good reason now to believe that we could have the enough
simple and adequate method of microobjects structure representation
in the frame of the ``graph geometry'' i.~e. beyond the
conventional space-time.

It is necessarily to keep in view that the basic ways of fundamental
investigations of the continuous manifolds with well-known
arbitrariness in the choice of metric and the various procedures of
``measurements'' are responsible all over for the mainstream of
physical research.

In this connection it is expediently to mark that we have as a rule the
different ``dimetrics'' where an initial element is the unit square
what allows to introduce an imaginary unity ``$i$'' and a
corresponding spectrum of numbers. However in the case of the
possible ``trimetrics'' where an initial element is the unit volume
we have a new spectrum of numbers without an imaginary unity
``$i$'' (for such uncommon ``trimetric'' there exists a specific
equality 3$^{3}$+4$^{3}$+5$^{3}$=6$^{3}$ analogous to the ``Egyptian
triangle'' equality 3$^{2}$+4$^{2}$=5$^{2}$). Further we may have
the different ``tetrametrics'' and so on.

\section*{2 Two-layer matrix scheme \\in the graph kinematics}

In part~I (see Ref.~[1]) it is carried out a study of the solution
scheme for an introduced physical graph kinematics which give rise
to the two-layer matrix presentation of the structure and the
scattering experiment of an arbitrary physical microobject within
the framework of the self-consistent ``graph geometry''. Below we
reproduce the typical examples of the application of this scheme to
the simple macro- and microobjects as a starting point for the
following analysis.

\pagebreak
\subsection*{2.1 Two-layer matrix description \\of the electric networks}

The use of two fundamental linear Kirchhoff's laws [4] for finding
of the branch (wire) currents $J$ of an electric network as a some
macroobject leads directly to the formation of two-layer, depending
on the resistances $\omega$, square matrix
$$\displaystyle \mathop{\bf M}_{(n\times
n)}(\omega)=\left\{\begin{array}{l}
\displaystyle \mathop{{\bf I}(\epsilon)}_{(v-1\times n)} \\
\displaystyle \mathop{{\bf C}(\epsilon)}_{(l\times
n)}\displaystyle \mathop{{\bf D}_{n}}_{(n\times n)}\!\!(\omega)
\end{array}\right\}\eqno(1)$$

\noindent
where the incidence matrix $\displaystyle
\mathop{{\bf I}(\epsilon)}_{(v-1\times n)}$ reflects the natural
``graph geometry'' of ($v,\,n$)-network with the skeleton trees
basis (see Ref.~[7]) and the independent loop matrix\linebreak
$\displaystyle\mathop{{\bf C}(\epsilon)}_{(l\times n)}\displaystyle
\mathop{{\bf D}_{n}}_{(n\times n)}\!\!(\omega)$ ($\displaystyle
\mathop{{\bf C}(\epsilon)}_{(l\times n)}$~--- cyclic matrix,
$\displaystyle \mathop{{\bf D}_{n}}_{(n\times n)}\!\!(\omega)$~---
diagonal matrix) characterizes the ``equilibrium conditions'' along
the $l=n-v+1$ loops of ($v,\,n$)-network with the same basis.

For a concrete network's graph, for example in the case of the full
graph $K_{\rm 4}$ with $\displaystyle v=4,\
n=\left({v\atop 2}\right)=6$, there exists as a reference frame the
set of 16 skeleton ($v=4,\ n=3$)~--- trees corresponding to
($3\times 3$)-submatrices of the incidence matrix
$\displaystyle\rm \mathop{{\bf I}\,(\epsilon)}_{(3\times 6)}^{}$ with
non-zero determinants. This set of skeleton ($v=4,\ n=3$)~---
trees constructs the form of denominator and numerators of the
resulting solution $ J\,(\omega,{\cal E})$ ({$\cal E$}~---
electromotive forces).

\subsection*{2.2 Two-layer matrix approximation \\ in the
physical graph kinematics}

In the case of a possible transition from the initial $S$-matrix
theory of Heisenberg [5] with Feynman integrals and corresponding
Feynman diagrams to the physical graph kinematics formalism, for
microobjects already, one takes place also the application of
two-layer matrix approximation with the analogous construction of
square matrix of (1) type. This approximation allows to determine
the graph internal 4-momenta $q$ on the mass shell $\displaystyle
(q^{2}=m^{2})$ by means of the adequate linear conditions of an
``extremal equilibrium'' in all vertices and along independent
loops of the physical graph with corresponding square matrix
$$\displaystyle \mathop{{\bf M}\,(\alpha)}_{(n\times
n)}^{}=\left\{\begin{array}{cc}
\displaystyle \mathop{{\bf I}\,(\epsilon)}_{(v-1\times n)}^{} \\
\displaystyle \mathop{{\bf C}\,(\epsilon)}_{(l\times
n)}^{}\displaystyle\mathop{{\bf D}_{n}(\alpha)}_{(n\times n)}^{}
\end{array}\right\};\eqno(2)$$

\noindent
here the ``upper layer''~--- incidence matrix $\displaystyle
\mathop{\rm {\bf I}\,(\epsilon)}_{(v-1\times n)}^{}$ factually
reflects again the natural ``graph geometry'' of physical
$(v,n)$-graph but with the root trees basis (see Ref. [7]) and
the ``under layer''~--- independent loop matrix $\displaystyle
\mathop{{\bf C}\,(\epsilon)}_{(l\times
n)}^{}\mathop{{\bf D}_{n}(\alpha)}_{(n\times n)}^{}$, depending on
the Feynman $\alpha$-parameters, determines already the extremal
conditions along the $l=n-v+1$ loops of the physical $(v,n)$-graph
with the same basis. The resulting $q\,(p,\alpha)$ solution
($p$~--- external 4-momenta of the discrete stable physical
microobjects) includes the non-zero determinants in denominator
and numerators which are responded to the set of appropriate root
trees.

In the next two sections the above-described physical graph
kinematics with two-layer matrices and root trees basis is
extended to a concrete analysis of the general properties of
different atomic, nuclear and subnuclear microobjects.

\section*{3 Root trees representation of the atomic \\ and
nuclear systems}

The essential characteristics of the atomic and nuclear shell
systems in the first iteration at least are to be extracted from
the main properties of the root $v$-trees. As it being known
already according to the Table 1 of part I (Ref. [1]) for this task
is to be considered the following interval:
$$\displaystyle 1\leqslant T_{v}\leqslant 115 \ \,(1 < v\leqslant
8).\eqno(3)$$

\noindent
Here is omitted the value $v=1$ as a trivial case corresponding to
an isolate vertex and we begin an investigation of the root trees
sequence in the indicated interval (3) with $v=2,\ T_{v=2}=1$ which
is responsible to the primary ``core'' of an atomic or nuclear
shell system (including possibly the spin in a given graph
description). Such ``core'' is an elementary root tree with one
``free''~--- terminal~--- vertex $\displaystyle v_{F}=1\,({\rm deg}\,(v_{F})=1)$ or
a representation of the initial simplest physical microobject.

Further, by the suggesting graph approximation~--- planned in
2.2~--- for the atomic and nuclear shell systems and taking into
account the excluded ``core'' vertex (or vertices) we can
identify the number of lines~--- ``edges''~--- for any root
$v$-tree, equal to $v-1$, with the principal quantum number ($QN$)
plus 1 (one ``core'' vertex here for the root tree in the form of a
path or a simple trail from the root vertex) $$\displaystyle
v-1=n_{QN}+1.$$

\noindent
Indeed, more exactly, if we denote through $\displaystyle
(v_{R}$)~--- the root vertex and through $\displaystyle
d\,(v_{R},v_{F})$~--- the distances between the single root vertex
$\displaystyle (v_{R})$ and the different ``free'' vertices
$\displaystyle (v_{F}$) (with $\displaystyle {\rm deg}\,(v_{F})=1)$
it is naturally to make the following evident assumption for the
values of principal $QN$
$$\displaystyle n_{QN}=d_{\max}(v_{R},v^{\circ}_{F})-1=v-2\eqno(4)$$

\noindent
and as usually the possible values for the orbital $QN$
$$\displaystyle l_{QN}:0,1,\ldots{}
,(n_{QN}-2)(n_{QN}-1)\eqno(4^{\prime})$$

\noindent
what corresponds to the number of ``free'' vertices $\displaystyle
(v_{F})$ with lower values of $\displaystyle d\,(v_{R},\, v_{F})$
but without of the ``core'' vertices $\displaystyle (v^{c}_{F})$
with $\displaystyle d\,(v_{R},v^{c}_{F})=1$.

Thus according to (4) the principal $QN$ is responsible for one
root $v$-tree from the full set~--- $\displaystyle T_{v}$~--- of
the root $v$-trees; namely, this root $v$-tree has the form of a
path (simple trail) from the root vertex $\displaystyle (v_{R})$
with the maximum distance $\displaystyle
d_{\max}(v_{R},v^{\circ}_{F})$ to a single ``free'' vertex denoted through
$\displaystyle (v^{\circ}_{F}):v^{\circ}_{F}=1$; the latter is
responsible for $\displaystyle l_{QN}=0$ at a given $\displaystyle
n_{QN}$. To the ``core'' vertices $\displaystyle (v^{c})$,
including ``free'' vertices $\displaystyle (v^{c}_{F})$, belong all
vertices with minimum distance $\displaystyle d\,(v_{R},v^{c})=1$
from the root vertex; these vertices are unconnected with the
representation of various $QN$ for shell systems.  At last the
``free'' vertices $\displaystyle (v_{F})$ with the intermediate
distances from the root vertex $\displaystyle (v_{R})$
$$\displaystyle 1 < d\,(v_{R},v_{F}) <
d_{\max}(v_{R},v^{\circ}_{F})$$

\noindent
are responsible for the rest values of $\displaystyle l_{QN} > 0$
from (4$^{\rm \prime}$). However in connection with the
above-stated graph approximation model it is reasonable to note
that beginning from $v\geqslant 5\,(n_{QN}\geqslant 3)$ the set of
``free'' vertices $\displaystyle (v_{F})$ of a concrete root tree,
of course with $\displaystyle d\,(v_{R},v_{F}) <
d_{\max}(v_{R},v^{\circ}_{F})=v-1$, may be associated at the same
values of $\displaystyle l_{QN}\ {\rm from}\ (4^{\prime})$.

Table 1 below includes, in agreement with an ``extremal
equilibrium'' consideration in subsection 2.2 (stable systems), the
rules of a filling of the electron shells in atoms $\displaystyle
(2n^{2}_{QN})$ and the nucleon shells in nuclei (magic numbers for
protons) in comparison with the parameters of root trees scheme
($v$, and in a hidden form $$\displaystyle
v_{R},v_{F},v^{\circ}_{F},v^{c},\ldots{}
,d_{\max}(v_{R},v^{\circ}_{F}),d\,(v_{R},v_{F}),$$ etc.) or straight
with the root $v$-trees numbers $\displaystyle (T_{v})$:

{\tabcolsep=8dd
\begin{center}
\newcommand{\Sloppy}{\emergencystretch 3em \tolerance 9999 }
\let\sloppy=\Sloppy
\begin{tabular}{c|c|c|c}
\multicolumn{4}{r}{\bf Table 1.}\\
\multicolumn{4}{c}{}\\[-7dd]
\hline
\multicolumn{1}{c|}{\rule{15dd}{0dd}} &
\multicolumn{1}{c|}{} &
\multicolumn{1}{c|}{} &
\multicolumn{1}{c}{} \\[-7dd]
 &
 &
Electron shells: &
Nucleon shells: \\
$v$ &
$T_{ v}$ &
2\,($v$--2)$^{\rm 2}$=2$n^{\rm 2}_{ QN}$ &
magic numbers \\
 &
 &
 &
for protons \\
 &
 &
 &
 \\[-6dd]
\hline
&&&                                                              \\
2                                                                 &
1                                                                 &
---                                                               &
---                                                              \\
&&&\\[-6dd]
\hline
&&&\\[-6dd]
3                                                                  &
2                                                                  &
2                                                                  &
2                                                                 \\
&&&\\[-6dd]
\hline
&&&\\[-6dd]
4                                                                  &
4                                                                  &
8                                                                  &
8                                                                 \\
&&&\\[-6dd]
\hline
&&&\\[-6dd]
5                                                                  &
9                                                                  &
18                                                                 &
                                                                  \\
&&&\\[-6dd]
\hline
&&&\\[-6dd]
6                                                                  &
20                                                                 &
32                                                                 &
20                                                                \\
&&&\\[-6dd]
\hline
&&&\\[-6dd]
                                                                   &
                                                                   &
                                                                   &
28                                                                \\
&&&\\[-6dd]
\hline
&&&\\[-6dd]
7                                                                  &
48                                                                 &
50                                                                 &
50                                                                \\
&&&\\[-6dd]
\hline
&&&\\[-6dd]
                                                                   &
                                                                   &
                                                                   &
82                                                                \\
&&&\\[-6dd]
\hline
&&&\\[-6dd]
8                                                                  &
115                                                                &
                                                                   &
114                                                               \\
\end{tabular}
\end{center} }

\pagebreak
\section*{4 Root trees evaluation of the subnuclear\\
microobject masses}

\vskip 4dd

It was shown in part I (see Ref. [1]) that a number of the root
$v$-trees $T_{v}$ is to satisfy to the next inequalities for any
$v>1$

$$\displaystyle T_{v+1}\geqslant 2 T_{v},\ T_{v+1} < 3 T_{v}.
\eqno(5)$$

For the sake of a graph description of the quasi-doublet or
quasi-triplet physical microobject constitution it is necessary to produce
from (5) the following chain of the recurrent inequalities

$$\displaystyle T_{v}\geqslant 2^{r}T_{v-r},\ T_{v} < 3 ^{s}T_{v-s}
\eqno(6)$$

\noindent
where $r,\, s=1,\,2,\,3,\ldots{} $ and so on, what allows to
compile the Table 2 (see below) for comparison of $T_{v}$ values at
$v\geqslant 9$ and their ``double- and triple-splitting
fragments'' with the experimental masses of appropriate subnuclear
microobjects (part I (Ref. [1]) contains only two illustrative
examples).

{\tabcolsep=2dd
\begin{center}
\newcommand{\Sloppy}{\emergencystretch 3em \tolerance 9999 }
\let\sloppy=\Sloppy
\begin{tabular}{c|c|c|l}
\multicolumn{4}{l}{{\bf Table 2.}  The fractional
``fragments'' of the root
trees
numbers $T_{v}$ ($v$ is the number}\\
\multicolumn{4}{l}{ of vertices) in comparison with
the experimental masses of ``suitable'' {\bf X}~--- microobjects}\\
\multicolumn{4}{l}{(see Ref. [6]) where an average error
of subnuclear particles
masses ranges 0 to 10 per cent. }\\
\hline
\multicolumn{1}{c|}{}                                        &
\multicolumn{1}{|c|}{}                                       &
\multicolumn{1}{|c|}{}                                       &
\multicolumn{1}{|c}{}                                       \\
\multicolumn{1}{c|}{$v$}                                     &
\multicolumn{1}{|c|}{$r,\, s$}                               &
\multicolumn{1}{|c|}{$T_{v}/2^{r}3^{s}$}                      &
\multicolumn{1}{|c}{$(M/m_{e})_{\exp}[{\bf X}]$}             \\
\multicolumn{1}{c|}{}                                        &
\multicolumn{1}{|c|}{}                                       &
\multicolumn{1}{|c|}{}                                       &
\multicolumn{1}{|c}{}                                       \\
\hline
\rule{5mm}{0.0dd}                                           &
\rule{9mm}{0.0dd}                                           &
\rule{17mm}{0.0dd}                                          &
\rule{95mm}{0.0dd}                                          \\
9&0,0&286&273$[{\bf \pi}^{\pm}]$,\, 264 $[{\bf \pi}^{\circ}]$\\
&1,1&47,7&$(T_{p=7})$                                       \\
&5,0&8,9&$(T_{p=5})$                                        \\
&&&                                                    \\[-5dd]
\hline
&&&                                                    \\[-5dd]
10&1,0&360&391[{\bf s}]                                    \\
&0,1&240&264 [${\bf \pi}^{\circ}$]                          \\
&1,1&119,8&$(T_{p=8})$                                     \\
&&&                                                    \\[-5dd]
\hline
&&&\\[-5dd]
11&0,0&1842&1875[${\bf \eta^{\prime} (958)}$],\, 1918
[${\bf f_{o}(980)}$],                                    \\
&&&1925 [${\bf a_{o} (980)}$],\, 1996 [${\bf \phi (1020)}$],\\
&&&1746 [${\bf K^{*} (892)}$]                              \\
&&&1839 [${\bf n}$],\, 1836 [${\bf p}$]               \\
&1,0&921&1071 [${\bf \eta}$],\, 966 [${\bf K^{\pm}}$],\, 974 [${\bf
K}^{\circ},\,\bar{{\bf K}}^{\circ}$]                       \\
&0,2&205&207 [${\bf \mu}$]                                 \\
&&&                                                    \\[-5dd]
\hline
&&&                                                    \\[-5dd]
12&0,0&4766&4495 [${\bf f_{2} (2300)}$],\, 4577 [${\bf f_{2}
(2340)}$]                                            \\
&&&4741 [${\bf D_{1} (2420)^{\circ}}$],\, 4962 [${\bf D_{s1}
(2536)^{\pm}}$]                                          \\
&&&4810 [${\bf D_{2}^{* \circ \pm} (2460)}$],\, 5036 [${\bf D_{SJ}
(2573)^{\pm}}$]                                           \\
&&& 4344 [${\bf N(2220)\cdot H_{19}}$],\, 4403 [${\bf N(2250)\cdot
G_{19}}$]                                                 \\
\end{tabular}
\end{center} }

{\tabcolsep=2dd
\begin{center}
\newcommand{\Sloppy}{\emergencystretch 3em \tolerance 9999 }
\let\sloppy=\Sloppy
\begin{tabular}{c|c|c|l}
\hline
\multicolumn{1}{c|}{}                                        &
\multicolumn{1}{|c|}{}                                       &
\multicolumn{1}{|c|}{}                                       &
\multicolumn{1}{|c}{}                                       \\
\multicolumn{1}{c|}{$v$}                                     &
\multicolumn{1}{|c|}{$r,\, s$}                               &
\multicolumn{1}{|c|}{$T_{v}/2^{r}3^{s}$}                      &
\multicolumn{1}{|c}{$(M/m_{e})_{\exp}[{\bf X}]$}             \\
\multicolumn{1}{c|}{}                                        &
\multicolumn{1}{|c|}{}                                       &
\multicolumn{1}{|c|}{}                                       &
\multicolumn{1}{|c}{}                                       \\
\hline
\rule{5mm}{0.0dd}                                           &
\rule{9mm}{0.0dd}                                           &
\rule{17mm}{0.0dd}                                          &
\rule{95mm}{0.0dd}                                          \\
&&&5088 [${\bf N(2600)\cdot I_{1,11}}$],\, 4736 [${\bf \Delta
(2420)\cdot H_{3,11}}$]                                   \\
&&&4599 [${\bf \Lambda (2340)\cdot H_{09}}$],\, 4403 [${\bf \Sigma
(2250)}$]                                                 \\
&&&4407 [${\bf \Omega (2250)^{-}}$],\, 4471 [${\bf
\Lambda_{c}^{+}}$]                                        \\
&&&5076 [${\bf \Lambda_{c} (2393)^{+}}$],\, 5140 [${\bf \Lambda_{c}
(2625)^{+}}$]                                             \\
&&&4804 [${\bf \Sigma_{c} (2455)}$],\, 4825 [${\bf \Xi_{c}^{+}}$]\\
&&&4834 [${\bf \Xi_{c}^{\circ}}$],\, 5174 [${\bf \Xi_{c}
(2645)}$]                                               \\
&&&5292 [${\bf \Omega_{c}^{\circ}}$]                     \\
&1,0&2383&2407 [${\bf a_{1} (1260)}$],\, 2409 [${\bf b_{1}
(1235)}$]                                           \\
&&&2290 [${\bf h_{1} (1170)}$],\, 2495 [${\bf f_{2} (1270)}$]\\
&&&2508 [${\bf f_{1} (1285)}$],\, 2534 [${\bf \eta (1295)}$]\\
&&&2544 [${\bf \pi (1300)}$],\, 2580 [${\bf a_{2}(1320)}$]\\
&&&2491 [${\bf K_{1}(1270)}$], 2183 [${\bf \Lambda^{\circ}}$]                            \\
&&&2328 [${\bf \Sigma^{+}}$],\, 2334 [${\bf \Sigma^{\circ}}$],\,
2343 [${\bf \Sigma^{-}}$]                                \\
&&&2573 [${\bf \Xi^{\circ}}$],\, 2585 [${\bf \Xi^{-}}$], 2544 [{\bf c}] \\
&2,1&397&391 [${\bf s}$]                                 \\
&1,2&265& 264 [${\bf \pi^{\circ}}$]                      \\
&2,3&44& ($T_{p=7}$)                                     \\
&&&                                                    \\[-5dd]
\hline
&&&                                                    \\[-5dd]
13&1,0&6243&5829 [${\bf \eta_{c}(1S)}$],\, 6060 [${\bf J/\psi}$]\\
&&&6683 [${\bf \chi_{co}(1P)}$],\, 6870 [${\bf \chi_{c1}(1P)}$]\\
&0,1&4162&3935 [${\bf f_{2} (2010)}$],\, 4000 [${\bf f_{4}
(2050)}$]                                                \\
&&&4495 [${\bf f_{2} (2300)}$],\, 4577 [${\bf f_{2}
(2340)}$]                                               \\
&&&4002 [${\bf K_{4}^{*} (2045)}$]                     \\
\end{tabular}
\end{center} }

{\tabcolsep=2dd
\begin{center}
\newcommand{\Sloppy}{\emergencystretch 3em \tolerance 9999 }
\let\sloppy=\Sloppy
\begin{tabular}{c|c|c|l}
\hline
\multicolumn{1}{c|}{}                                        &
\multicolumn{1}{|c|}{}                                       &
\multicolumn{1}{|c|}{}                                       &
\multicolumn{1}{|c}{}                                       \\
\multicolumn{1}{c|}{$v$}                                     &
\multicolumn{1}{|c|}{$r,\, s$}                               &
\multicolumn{1}{|c|}{$T_{v}/2^{r}3^{s}$}                      &
\multicolumn{1}{|c}{$(M/m_{e})_{\exp}[{\bf X}]$}             \\
\multicolumn{1}{c|}{}                                        &
\multicolumn{1}{|c|}{}                                       &
\multicolumn{1}{|c|}{}                                       &
\multicolumn{1}{|c}{}                                       \\
\hline
\rule{5mm}{0.0dd}                                           &
\rule{9mm}{0.0dd}                                           &
\rule{17mm}{0.0dd}                                          &
\rule{95mm}{0.0dd}                                          \\
&&&3927 [${\bf D^{*} (2007)^{\circ}}$],\, 3933 [${\bf D^{*}
(2010)^{\pm}}$]                                               \\
&&&3852 [${\bf D_{s}^{\pm}}$],\, 4129 [${\bf D^{* \, \pm}_{s}}$]\\
&&&4286 [${\bf N (2190)\cdot G_{17}}$],\, 4344 [${\bf N (2220)\cdot
H_{19}}$]                                                  \\
&&&4403 [${\bf N (2250) \cdot G_{19}}$,\, 3816 [${\bf \Delta
(1950)\cdot F_{37}}$]                                      \\
&&&4110 [${\bf \Lambda (2100) \cdot G_{07}}$],\, 4129 [${\bf
\Lambda (2110)\cdot F_{05}}$]                              \\
&&&4599 [${\bf \Lambda (2350) \cdot H_{09}}$],\, 3796 [${\bf \Sigma
(1940) \cdot D_{13}}$]                                     \\
&&&3972 [${\bf \Sigma (2030) \cdot F_{17}}$],\, 4403 [${\bf \Sigma
(2250)}$]                                                   \\
&&&3816 [${\bf \Xi (1950)}$],\, 3963 [${\bf \Xi (2030)}$]   \\
&&&4471 [${\bf \Lambda_{c}^{+}}$],\, 4407 [${\bf \Omega
(2250)^{-}}$]                                              \\
&2,0&3122&2867 [${\bf \rho (1450)}$],\, 2941 [${\bf f_{o}
(1500)}$]                                                 \\
&&&2959 [${\bf f_{1}(1510)}$],\, 2984 [${\bf f_{2}^{\prime}
(1525)}$]                                                \\
&&&3227 [${\bf \omega (1600)}$],\, 3264 [${\bf \omega_{3}
(1670)}$]                                               \\
&&&3268 [${\bf \pi_{2} (1670)}$],\, 3288 [${\bf \phi (1680)}$]\\
&&&3309 [${\bf \rho_{3} (1690)}$],\, 3327 [${\bf \rho (1700)}$]\\
&&&3321 [${\bf f_{J} (1710)}$],\, 3354 [${\bf K^{*} (1680)}$] \\
&&&3464 [${\bf K_{3}^{*} (1780)}$],\, 3470 [${\bf K_{2}
(1770)}$]                                                   \\
&&&2975 [${\bf N (1520)}$],\, 3004 [${\bf N (1530) \cdot
S_{11}}$],\, 3229 [${\bf N (1650) \cdot S_{11}}$]            \\
&&&3278 [${\bf N(1675) \cdot D_{15}}$],\, 3288 [${\bf N (1680)\cdot
F_{15}}$]                                                  \\
&&&3327 [${\bf N(1700) \cdot D_{13}}$],\, 3346 [${\bf N (1710)\cdot
P_{11}}$]                                                  \\
&&&3366 [${\bf N(1720) \cdot P_{13}}$],\, 3131 [${\bf \Delta
(1600)\cdot P_{33}}$]                                         \\
&&&3170 [${\bf \Delta (1620) \cdot S_{31}}$],\, 3327 [${\bf \Delta
(1700)\cdot D_{33}}$]                                         \\
&&&3131 [${\bf \Lambda (1600) \cdot P_{01}}$],\, 3268 [${\bf
\Lambda (1670)\cdot S_{01}}$]                                  \\
&&&3307 [${\bf \Lambda (1690) \cdot D_{03}}$],\, 2974 [${\bf
\Lambda (1520)\cdot D_{03}}$]                                   \\
\end{tabular}
\end{center} }

{\tabcolsep=2dd
\begin{center}
\newcommand{\Sloppy}{\emergencystretch 3em \tolerance 9999 }
\let\sloppy=\Sloppy
\begin{tabular}{c|c|c|l}
\hline
\multicolumn{1}{c|}{}                                        &
\multicolumn{1}{|c|}{}                                       &
\multicolumn{1}{|c|}{}                                       &
\multicolumn{1}{|c}{}                                       \\
\multicolumn{1}{c|}{$v$}                                     &
\multicolumn{1}{|c|}{$r,\, s$}                               &
\multicolumn{1}{|c|}{$T_{v}/2^{r}3^{s}$}                      &
\multicolumn{1}{|c}{$(M/m_{e})_{\exp}[{\bf X}]$}             \\
\multicolumn{1}{c|}{}                                        &
\multicolumn{1}{|c|}{}                                       &
\multicolumn{1}{|c|}{}                                       &
\multicolumn{1}{|c}{}                                       \\
\hline
\rule{5mm}{0.0dd}                                           &
\rule{9mm}{0.0dd}                                           &
\rule{17mm}{0.0dd}                                          &
\rule{95mm}{0.0dd}                                          \\
&&&3248 [${\bf \Sigma (1660) \cdot P_{11}}$],\, 3268 [${\bf
\Sigma (1670)\cdot D_{13}}$]                                     \\
&&&3425 [${\bf \Sigma (1750) \cdot S_{11}}$],\, 2994 [${\bf
\Xi (1530)\cdot P_{13}}$]                                        \\
&&&3307 [${\bf \Xi (1690)}$], \, 3273 [${\bf \Omega^{-}}$]       \\
&1,1&2081&1918 [${\bf f_{0} (980)}$],\, 1925 [${\bf a_{0}
(980)}$]                                                       \\
&&&1996 [${\bf \phi (1020)}$],\, 2290 [${\bf h_{1} (1170)}$]    \\
&&& 2183 [${\bf \Lambda^{\circ}}$]                             \\
&0,2&1387&1507 [${\bf \rho (770)}$],\, 1530 [${\bf \omega
(792)}$]                                                     \\
&2,1&1040&1071 [${\bf \eta}$]                                \\
&&&966 [${\bf K^{\pm}}$],\, 974 [${\bf K^{\circ},\,
\bar{K}^{\circ}}$]                                         \\
&&&                                                    \\[-5dd]
\hline
&&&                                                    \\[-5dd]
14&0,1& 10991 &10330
[${\bf B^{\pm}}$],\, 10331 [${\bf B^{\circ}}$],\,
10420 [${\bf B^{*}}$]                      \\
&&& 10519 [${\bf B^{\circ}_{s}}$],\, 11039 [${\bf
\Lambda^{\circ}_{b}}$]                            \\
&2,0& 8243 &8415 [{\bf b}], \, 7906 [${\bf \psi (4040)}$], \,
8139 [${\bf \psi (4160)}$],\, 8640 [${\bf \psi (4415)}$] \\
&0,2& 3664 &3474 [${\bf \tau}$], \, 3628 [${\bf \phi_{3} (1850)}$],
\, 3935 [${\bf f_{2} (2010)}$]     \\
&&&4000 [${\bf f_{4} (2050)}$],\, 3553 [${\bf K_{2}
(1820)}$] \, 3354 [${\bf K^{*} (1680)}$]             \\
&&&3464 [${\bf K_{3}^{*} (1780)}$],\,
3470 [${\bf K_{2} (1770)}$],\,
4002 [${\bf K_{4}^{*} (2045)}$]                     \\
&&&3649 [${\bf D^{\circ}}$],\, 3658 [${\bf D^{\pm}}$],\,
3927 [${\bf D^{*} (2007)^{\circ}}$]                             \\
&&&3933 [${\bf D^{*} (2010)^{\pm}}$],\,
3852 [${\bf D^{\pm}_{s}}$]              \\
&&&3346 [${\bf N (1710)\cdot P_{11}}$],\, 3366 [${\bf N
(1720)\cdot P_{13}}$],\,  3718 [${\bf \Delta (1900)
\cdot S_{31}}$]               \\
&&&3728 [${\bf \Delta (1905) \cdot F_{35}}$],\,
3738 [${\bf \Delta (1910)\cdot P_{31}}$],\,
3757 [${\bf \Delta (1920)\cdot P_{33}}$]              \\
&&&3777 [${\bf \Delta (1930) \cdot D_{35}}$],\,
3816 [${\bf \Delta (1950)\cdot F_{37}}$],\,
3523 [${\bf \Lambda (1800)\cdot S_{01}}$]              \\
&&&3544 [${\bf \Lambda (1810) \cdot P_{01}}$],\,
3562 [${\bf
\Lambda (1820)\cdot F_{05}}$],\,
3581 [${\bf \Lambda (1830) \cdot D_{05}}$]            \\
&&&3699 [${\bf \Lambda (1890) \cdot P_{03}}$],\,
3425 [${\bf \Sigma
(1750) \cdot S_{11}}$],\,
3474 [${\bf \Sigma (1775)
\cdot D_{15}}$]                                     \\
\end{tabular}
\end{center} }

{\tabcolsep=2dd
\begin{center}
\newcommand{\Sloppy}{\emergencystretch 3em \tolerance 9999 }
\let\sloppy=\Sloppy
\begin{tabular}{c|c|c|l}
\hline
\multicolumn{1}{c|}{}                                        &
\multicolumn{1}{|c|}{}                                       &
\multicolumn{1}{|c|}{}                                       &
\multicolumn{1}{|c}{}                                       \\
\multicolumn{1}{c|}{$v$}                                     &
\multicolumn{1}{|c|}{$r,\, s$}                               &
\multicolumn{1}{|c|}{$T_{v}/2^{r}3^{s}$}                      &
\multicolumn{1}{|c}{$(M/m_{e})_{\exp}[{\bf X}]$}             \\
\multicolumn{1}{c|}{}                                        &
\multicolumn{1}{|c|}{}                                       &
\multicolumn{1}{|c|}{}                                       &
\multicolumn{1}{|c}{}                                       \\
\hline
\rule{5mm}{0.0dd}                                           &
\rule{9mm}{0.0dd}                                           &
\rule{17mm}{0.0dd}                                          &
\rule{95mm}{0.0dd}                                          \\
&&&3748 [${\bf \Sigma (1915) \cdot F_{15}}$],\, 3796 [${\bf \Sigma
(1940) \cdot D_{13}}$], \,
3972 [${\bf \Sigma (2030) \cdot F_{17}}$]             \\
&&&3668 [${\bf \Xi (1820) \cdot D_{13}}$],\,
3816 [${\bf \Xi (1950)}$], \, 3963
[${\bf \Xi (2030)}$]                   \\
&2,1&2748 &2544 [{\bf c}],\,
2508 [${\bf f_{1} (1285)}$],\, 2534 [${\bf \eta
(1295)}$]                                                 \\
&&&2544 [${\bf \pi(1300)}$],\, 2580 [${\bf a_{2}
(1320)}$],\,
2681 [${\bf f_{0} (1370)}$]                          \\
&&&2769 [${\bf \eta (1440)}$],\, 2777 [${\bf \omega
(1420)}$],\,
2792 [${\bf f_{1} (1420)}$]                    \\
&&&2867 [${\bf \rho (1450)}$],\, 2941 [${\bf f_{0} (1500)}$],\,
2959 [${\bf f_{1} (1510)}$]          \\
&&&2984 [${\bf f_{2}' (1525)}$],\, 2744 [${\bf K_{1} (1400)}$],\,
2763 [${\bf K^{*} (1410)}$]                          \\
&&&2796 [${\bf K_{0}^{*} (1430)}$],\, 2798 [${\bf K^{*}_{2}
(1430)}$]                          \\
&&&2818 [${\bf N (1440) \cdot P_{11}}$],\, 2975 [${\bf
N (1520) \cdot D_{13}}$],\,  3004 [${\bf N
(1535) \cdot S_{11}}$]                 \\
&&&2753 [${\bf \Lambda (1405) \cdot S_{01}}$],\, 2974 [${\bf
\Lambda (1520)\cdot D_{03}}$],\,
2710 [${\bf \Sigma (1385) \cdot
P_{13}}$]                                   \\
&&&2573 [${\bf \Xi^{\circ} }$],\, 2585 [${\bf
\Xi^{-} }$],\,
2994 [${\bf \Xi (1530) \cdot
P_{13}}$]                                   \\
&1,2&1832&1875 [${\bf \eta' (958) }$],\, 1918 [${\bf
f_{0} (980)}$],\,
1925 [${\bf a_{0} (980)}$]                                   \\
&&&1996 [${\bf \phi (1020)}$],\, 1746 [${\bf K^{*}
(892)}$]                      \\
&&&1836 [${\bf p}$],\, 1839 [${\bf n}$]               \\
&&& ($T_{p=11}$)                               \\
&&&                                                    \\[-5dd]
\hline
&&&                                                    \\[-5dd]
15&2,1& 7318&6683 [${\bf \chi_{c0} (1P)}$],\, 6870
 [${\bf \chi_{c1}  (1P)}$], \,
 6959 [${\bf \chi_{c2} (1P)}$] \\
&&&7213 [${\bf \psi (2S)}$],\, 7377
 [${\bf \psi  (3770)}$], \,
 7906 [${\bf \psi (4040)}$] \\
&&&\\
\hline
16&2,1& 19615&18513 [${\bf \Upsilon (1S)}$],\, 19295
 [${\bf \chi_{b0}  (1P)}$], \,
 19358 [${\bf \chi_{b1} (1P)}$] \\
&&&19400 [${\bf \chi_{b2} (1P)}$],\, 19615
 [${\bf \Upsilon (2S)}$], \,
 20024 [${\bf \chi_{b0} (2P)}$] \\
&&&20069 [${\bf \chi_{b1} (2P)}$],\, 20095
 [${\bf \chi_{b2} (2P)}$], \,
 20265 [${\bf \Upsilon (3S)}$] \\
&&& 20704 [${\bf \Upsilon (4S)}$],\, 21262
 [${\bf \Upsilon (10870)}$], \,
 21564 [${\bf \Upsilon (11020)}$] \\
\end{tabular}
\end{center} }

{\tabcolsep=2dd
\begin{center}
\newcommand{\Sloppy}{\emergencystretch 3em \tolerance 9999 }
\let\sloppy=\Sloppy
\begin{tabular}{c|c|c|l}
\hline
\multicolumn{1}{c|}{}                                        &
\multicolumn{1}{|c|}{}                                       &
\multicolumn{1}{|c|}{}                                       &
\multicolumn{1}{|c}{}                                       \\
\multicolumn{1}{c|}{$v$}                                     &
\multicolumn{1}{|c|}{$r,\, s$}                               &
\multicolumn{1}{|c|}{$T_{v}/2^{r}3^{s}$}                      &
\multicolumn{1}{|c}{$(M/m_{e})_{\exp}[{\bf X}]$}             \\
\multicolumn{1}{c|}{}                                        &
\multicolumn{1}{|c|}{}                                       &
\multicolumn{1}{|c|}{}                                       &
\multicolumn{1}{|c}{}                                       \\
\hline
\rule{5mm}{0.0dd}                                           &
\rule{9mm}{0.0dd}                                           &
\rule{17mm}{0.0dd}                                          &
\rule{95mm}{0.0dd}                                          \\
&7,0&1839&1839 [${\bf n}$],\, 1836 [${\bf p}$],\ldots,\,
 ($T_{p=11}$)               \\
&11,0&115& ($T_{p=8}$)                               \\
\hline
17&1,0&317424&352250 [${\bf t}$]               \\
&2,0&158712& 157201 [${\bf W^{\pm}}$]
\\ &&&                                                    \\[-5dd]
\hline
&&&                                                    \\[-5dd]
18&0,2&191240&178448 [${\bf Z^{\circ}}$]               \\
\end{tabular}
\end{center} }

It is easy to examine that in a given iteration (an average error
ranges 0 to 10 per cent) factually all subnuclear masses from [6]
are presented in Table 2 (everywhere in Table 2, in accordance with Ref.
[7], number of vertices $p$ identical to $v$). The root trees numbers or
 its fractional
``fragments'' are fitted with indicated precision to experimental
values of all known particle masses. At the same time there take
place the ``mixing'' of meson and baryon particle categories and
the ``concentration'' of particle masses within an interval
$v$=9$\div$14.

\section*{5 Conclusions}

The above-developed, in section 2, two-layer matrix method for the
study and adequate description of any discrete microobject based on
the strongly derived from an incidence matrix $\displaystyle
\mathop{ {\bf I}\,(\epsilon)}_{(v-1\times n)}$ (upper layer) natural
``graph microgeometry'', generally unique for every concrete
microobject, allows to refuse from the common space-time treatment
owing to the appearance of an equivalent, to a certain degree,
graph language (a ``third language''). On the other hand, as it was
demonstrated earlier, the resulting solutions of the most equations
of the contemporary field theory after the corresponding
mathematical ``contractions'', in the direction to the particle~---
like solutions with a cardinal decrease of degrees of freedom,
could be represented as the systems of suitable diagrams or graphs
(of different types and interpretations).  Therefore due to
Riemann's general conception for the discrete manifolds (see Ref.
[2]) and Kirchhoff's method of skeleton trees for an electric
network (see Ref. [4]) we should analyse any micro as well as macro
discrete physical objects within their proper natural ``graph
geometry'' beyond space-time. In addition it is important to
emphasize that the ``graph geometry'' must be generated
straightforwardly from the discrete physical object characteristics
and probably for this reason it cann't serve as a frame
for physical reality in the form of theoretical
or experimental data as opposed to the continuous physical models
with ``external geometry''.

\end{document}